\documentclass[aps,prl,preprint,superscriptaddress]{revtex4-1}
\usepackage{lineno}
\usepackage[dvipdfmx]{graphicx}
\usepackage{dcolumn}
\usepackage{bm}
\usepackage{siunitx} 
\usepackage{color}

\begin{document}
\bibliographystyle{apsrev4-1}
\title{Polar State induced by Block-type Lattice Distortions in BaFe$_2$Se$_3$ with Quasi-One-Dimensional Ladder Structure}

\author{Takuya Aoyama}
 \email{aoyama@tohoku.ac.jp}
 \affiliation{%
  Department of Physics, Graduate School of Science, Tohoku University, 6-3, Aramaki Aza-Aoba, Aoba-ku, Sendai, Miyagi 980-8578, Japan.
 }%

\author{Satoshi Imaizumi}%
\affiliation{%
 Department of Physics, Graduate School of Science, Tohoku University, 6-3, Aramaki Aza-Aoba, Aoba-ku, Sendai, Miyagi 980-8578, Japan.
}%

\author{Takuya Togashi}%
\affiliation{%
 Department of Physics, Graduate School of Science, Tohoku University, 6-3, Aramaki Aza-Aoba, Aoba-ku, Sendai, Miyagi 980-8578, Japan.
}%

\author{Yoshifumi Sato}%
\affiliation{%
 Department of Physics, Graduate School of Science, Tohoku University, 6-3, Aramaki Aza-Aoba, Aoba-ku, Sendai, Miyagi 980-8578, Japan.
}%

\author{Kazuki Hashizume}%
\affiliation{%
 Department of Physics, Graduate School of Science, Tohoku University, 6-3, Aramaki Aza-Aoba, Aoba-ku, Sendai, Miyagi 980-8578, Japan.
}%

\author{Yusuke Nambu}
\affiliation{
 Institute for Materials Research, Tohoku University, Sendai, Miyagi 980-8577,Japan.
}%

\author{Yasuyuki Hirata}
\affiliation{%
 Institute for Solid State Physics, The University of Tokyo, Kashiwa, Chiba 277-8581, Japan.
}

\author{Masakazu Matsubara}
\affiliation{%
  Department of Physics, Graduate School of Science, Tohoku University, 6-3, Aramaki Aza-Aoba, Aoba-ku, Sendai, Miyagi 980-8578, Japan.
}

\author{Kenya Ohgushi}
\affiliation{%
 Department of Physics, Graduate School of Science, Tohoku University, 6-3, Aramaki Aza-Aoba, Aoba-ku, Sendai, Miyagi 980-8578, Japan.
}%

\date{\today}

\begin{abstract}
Temperature dependent crystal structures of the quasi-one-dimensional ladder material BaFe$_2$Se$_3$ are examined.
Combining the optical second harmonic generation (SHG) experiments and neutron diffraction measurements, we elucidate the crystal structure with $Pmn2_1$ space group in the low-temperature phase below $T_{s2}$ = 400 K, further above N\'{e}el temperature.
This low-temperature phase loses the spatial inversion symmetry, where a resultant macroscopic polarization emerges along the rung direction.
The transition is characterized by block-type lattice distortions with the magneto-striction mechanism.
Change in the electrical resistivity and the magnetic susceptibility across the polar-nonpolar transition also suggests a modification of the electronic states reflecting the structural instability.
Consistency and discrepancy with the existing theory are discussed.

\end{abstract}


\maketitle

The strong electron correlation effect is known to bear exotic quantum states of matter~\cite{Mott1990, Imada1998}.
Representative examples are the high-temperature superconductivity in cuprates and iron-based materials~\cite{Keimer2015,Hosono2015}, and the multiferroics with simultaneous presence of more than two ferroic order parameters~\cite{Schmid2008}.
Although both of the high-temperature superconductivity and the multiferroics share common scientific backgrounds~\cite{Liang2013,Kruger2009,Scagnoli696}, the relationship between these two states is rarely elucidated from both theoretical and experimental viewpoints.
Within this context, BaFe$_2$Se$_3$ is a nice platform, since it is theoretically predicted to be in a multiferroic state at the ambient pressure, and it is experimentally shown to become superconducting under pressure~\cite{Luo2013, Dong2014}.
The detailed study on BaFe$_2$Se$_3$ is considered to be a good starting point for exploring a fertile research field of the superconducting multiferroics~\cite{Kanasugi2018}.

The crystal structure of BaFe$_2$Se$_3$ is composed of edge-shared FeSe$_4$ tetrahedra forming a quasi-one-dimensional ladder structure of Fe atoms (Fig. 1(a)).
The structure can be regarded as the case one-third of the Fe atom stripes are removed from the two-dimensional square lattice of Fe atoms, reminding us of a close relationship with iron-based superconductors~\cite{Caron2011, Lei2011, Krzton-Maziopa2012,Lv2013}.
Superconductivity has been observed quite recently under high pressures in BaFe$_2$Se$_3$~\cite{Ying2017, Zhang2018}.
The material is frequently argued in comparison with a related material BaFe$_2$S$_3$, which also becomes superconducting under pressures~\cite{Takahashi2015, Yamauchi2015, Arita2015, Suzuki2015, Patel2016}; however these two compounds have a striking difference in the crystal structure.
Whereas BaFe$_2$S$_3$ has the $Cmcm$ space group in the whole temperature range measured (Fig. 1 (b)), BaFe$_2$Se$_3$ exhibits a structural phase transition at 660 K (= $T_{s1}$) from the high-temperature $Cmcm$ space group to low-temperature $Pnma$ space group~\cite{Svitlyk2013}, which induces small distortions both in intra and inter ladder structures (Fig. 1 (c)).
Below $T_{s1}$, Fe-Fe bonds along the leg direction become staggered in the anti-phase manner between the two adjacent legs, and the ladder planes are slightly rotated from the principal axes of the orthorhombic structure.
At around 400 K (= $T_{s2}$), the second structural transition occurs, which is evidenced by a shift of ($h00$) reflection in the X-ray diffraction profile and a sharp peak in the differential scanning calorimetry signal~\cite{Svitlyk2013}.
However, the nature of this structural transition is uncovered, which is the central question of this study.

With further cooling BaFe$_2$Se$_3$, an antiferromagnetic order develops below the N\'{e}el temperature ($T_{\rm N}$) of 220$\sim$255 K.
The magnetic structure was determined to be a block-type one with the magnetic easy axis along the layer direction, which is distinct from the stripe-type magnetic ordering with the magnetic easy axis of the rung direction in BaFe$_2$S$_3$~\cite{Luo2013,Lovesey2015}.
Block-type magnetically ordered phase is theoretically predicted to have large ferroelectric polarizations reflecting the broken inversion symmetry~\cite{Dong2014}.
The related important point is that the block-type magnetic structure within $Pnma$ involves basis functions over separate irreducible representations (irreps) to form the corepresentation~\cite{Nambu2012}, which puzzles us in the light of the second-order fashion of the magnetic transition.
Therefore, the actual crystal structure is expected to have lower symmetry than $Pnma$ even in the paramagnetic phase ($T_{\rm N}<T<T_{s2}$).

In this study, we performed the optical second harmonic generation (SHG) measurements for BaFe$_2$Se$_3$.
Together with re-analysis on powder neutron diffraction profiles, we show that the phase transition at 400 K is the structural transition from unpolar $Pnma$ to polar $Pmn2_1$ space group.
The transition is triggered by block-type lattice distortions through the spin-lattice coupling, and the resultant $Pmn2_1$ structure can compatibly accommodate the block-type magnetic structure.
Indeed linear spin-wave calculations with assuming this space group well account for observed magnetic excitations~\cite{Mourigal2015}.
Relationship between our experimental findings and electronic nematic transitions widely observed in the iron-based superconductors will be discussed.

High-quality single crystals of BaFe$_2$Se$_3$ were grown by the melt-growth method.
Stoichiometric amounts of elemental Ba shots, Fe powders, and Se powders in a carbon crucible were sealed into an evacuated quartz ampoule.
The ampoule was slowly heated up to 1373 K, kept for 48 hrs, and slowly cooled to room temperature.
The powder X-ray diffraction using the Cu-$K_{\alpha}$ radiation indicates no trace of impurity phases.
The electrical resistivity was measured by the standard four-probe method.
The magnetic susceptibility was collected by using a superconducting quantum interference device magnetometer with applying magnetic field along the $a$ (leg direction) and $b$ (layer direction) axes below room temperature, and only along the $a$ axis above room temperature.
For the optical measurements, the oriented crystals were polished by a sand-paper and Al$_2$O$_3$ fine powders.
The SHG experiments were performed in the experimental setup shown in Fig. 3(a).
Incident pulsed light with a wavelength of 800 nm (1.55 eV) and the pulse duration of 130 fs at a repetition rate of 1 kHz is generated by using a Ti:sapphire regenerative amplifier system, and is irradiated on the sample.
The laser power is $\sim$1 mW and the spot size is typically $\sim$200 $\mu$m.
The reflected light with twice energy is detected by a photomultiplier tube.
Optical measurements at low temperatures were performed by using He and N$_2$ flow cryostats.

\begin{figure}[h]
\centering
\includegraphics[width=7cm]{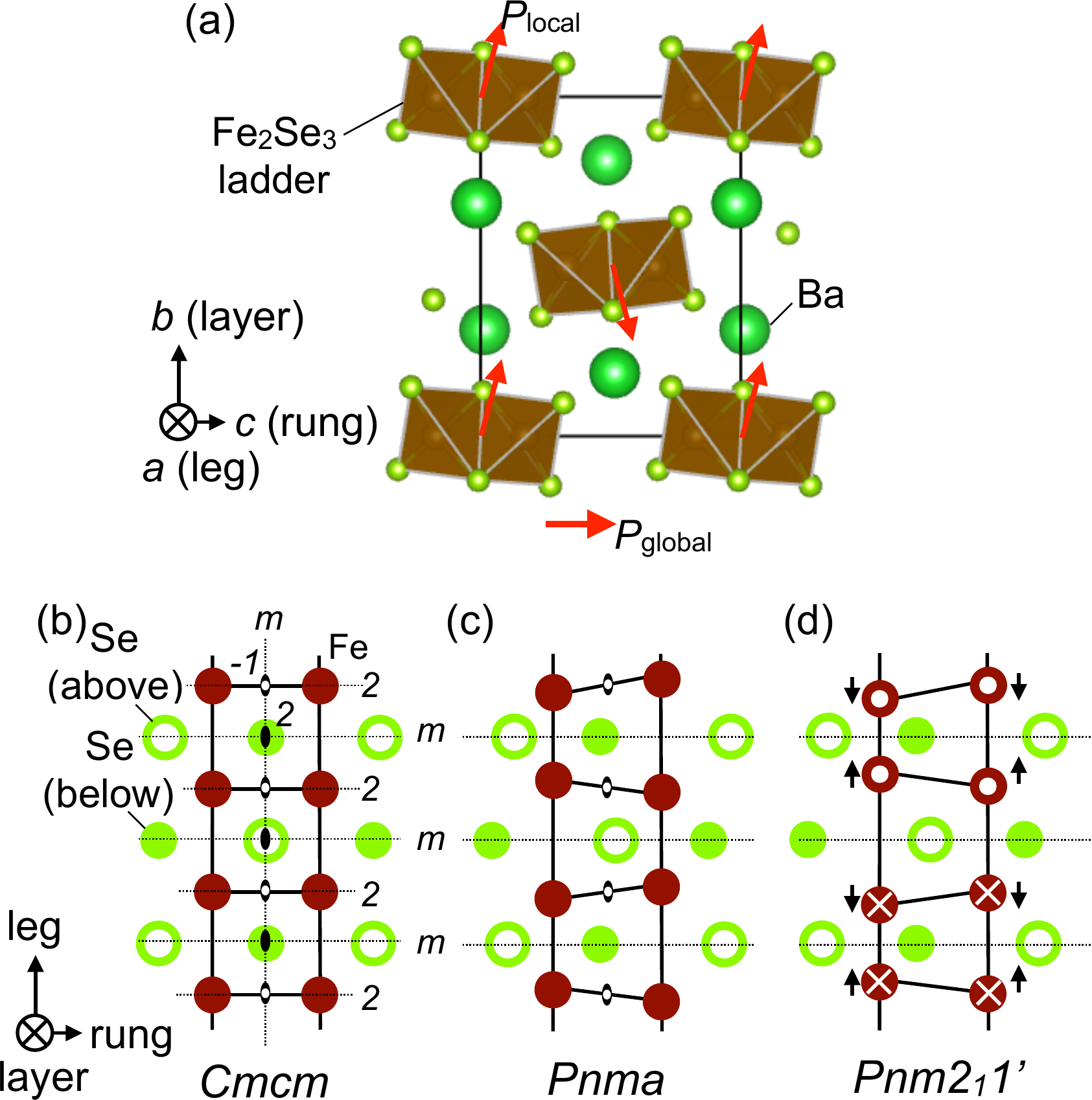}
\caption{(a) Crystal structure of BaFe$_2$Se$_3$ in the $Pmn2_1$ phase. The principle axes ($a$, $b$, and $c$) are defined under the $Pmn2_1$ notation. Red arrows indicate the electric polarization induced below $T_{s2}$. (b-d) Schematic drawings of the local ladder structure with the space group of (b) $Cmcm$, (c) $Pnma$, and (d) $Pmn2_11^{\prime}$. Preserved symmetry operations are also shown. The block-type magnetic structure is also shown in (d) with white circles (up spins) and crosses (down spins).}
\label{Fig1}
\end{figure}

\begin{figure}[h]
\centering
\includegraphics[width=7cm]{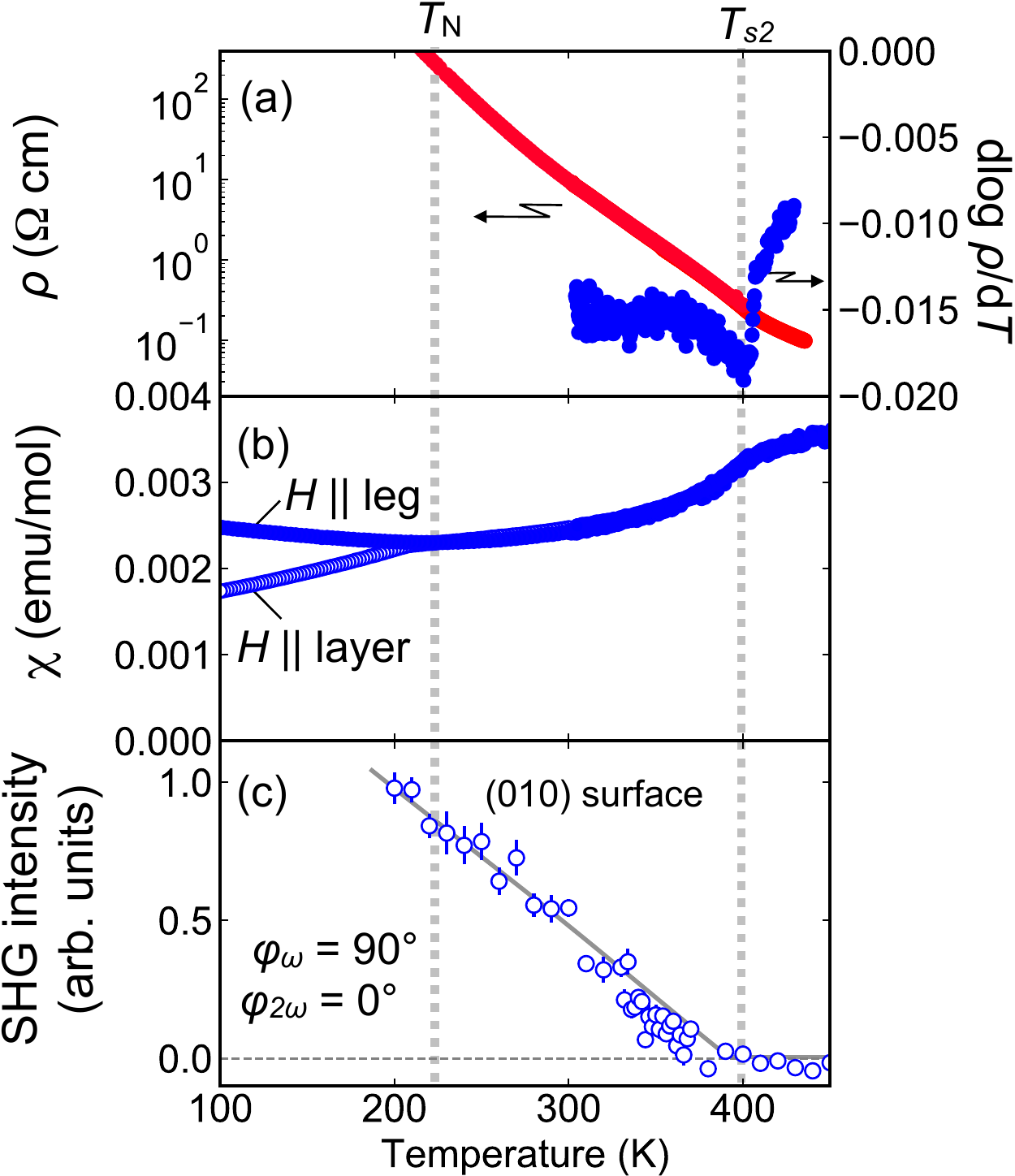}
\caption{Temperature dependence of physical properties of BaFe$_2$Se$_3$. (a) Electrical resistivity ($\rho$) with applying the current parallel to the leg direction (left), and the temperature derivative of log$\rho$ (right). (b) Magnetic susceptibility $(\chi)$ under the external magnetic field $(H)$ of 1 T applied along the layer (open) and leg (filled) directions. (c) Second harmonic generation (SHG) intensity in the experimental setup shown in Figs. 3(a) and 3(b). Below the structural transition temperature of $T_{s2} \sim $ 400 K, the SHG intensity due to the breaking of the spatial inversion symmetry develops.}
\label{Fig2}
\end{figure}

Figure 2 summarizes fundamental physical properties of BaFe$_2$Se$_3$.
The electrical resistivity ($\rho$) shown in Fig. 2(a) exhibits the Arrhenius-type temperature dependence with small excitation energy of 0.2 eV, which  is consistent with the energy gap observed in the optical conductivity spectra shown in Supplemental Materials~\cite{supple}.
The $\rho$ curve shows an anomaly at the second structural transition temperature of $T_{s2}$, as can be clearly seen from the dip-like feature in the temperature derivative of log$\rho$, indicating that the insulating behavior is enhanced in the low-temperature phase.
Temperature dependence of the magnetic susceptibility ($\chi$) taken under the magnetic field of 1 T is shown in Fig. 2(b).
Besides the reported anomaly corresponding to the antiferromagnetic transition at $T_{\rm N}$ = 220 K, we observe a clear anomaly at $T_{s2}$, suggesting a large change in electronic states is likely across $T_{s2}$.

To obtain further insights into the structural transition at $T_{s2}$, we performed optical SHG experiments.
The SHG is the second-order nonlinear optical phenomena, in which photons with twice the energy of initial photons are generated from samples.
The SHG intensity ($I$) is represented by $I(2\omega) \propto |\mu_0\frac{\partial ^2\vec{P}}{\partial t^2} + \mu_0 (\nabla \times \frac{\partial \vec{M}}{\partial t})|^2 $, which consists of an electric dipole contribution $P_i(2\omega) = \varepsilon_0\chi^{{\rm polar}}_{ijk} E_j(\omega)E_k(\omega)$ and a magnetic dipole contribution $M_i(2\omega) = \varepsilon_0c\chi^{{\rm axial}}_{ijk} E_j(\omega)E_k(\omega)$, where $P$, $M$, and $E$ stand for the electric polarization, magnetic dipole moment and electric field, respectively~\cite{Fiebig05}.
The tensors $\chi^{\rm polar}_{ijk}$ and $\chi^{\rm axial}_{ijk}$ are the third-rank SHG tensors having polar and axial characteristics, respectively.
The SHG tensors are governed by the point group of the system, and the particularly important is that polar tensors can be finite only in noncentrosymmetric systems.
For example, in the centrosymmetric $mmm$ symmetry (the point group of $Cmcm$ and $Pnma$ space groups), all the components in $\chi^{{\rm polar}}$ should be zero, whereas $\chi^{\rm axial}_{abc}$, $\chi^{\rm axial}_{bca}$, and $\chi^{\rm axial}_{cab}$ components of $\chi^{\rm axial}$ can be finite.

We first performed experiments in a configuration shown in Figs. 3(a) and 3(b): the incident laser is irradiated on the (010) surface and the reflected SHG light normal to the surface is detected; the polarization of the initial light ($\phi_{\omega}$) and that of the reflected SHG light ($\phi_{2\omega}$) were analyzed by linear polarizers.
The advantage of this setup is that magnetic-dipole contributions ($\chi^{\rm axial}_{abc}$, $\chi^{\rm axial}_{bca}$, and $\chi^{\rm axial}_{cab}$) expected for the $mmm$ point group are undetectable due to the symmetrical reason.
Nevertheless, as can be seen from Fig. 2(c), when $\phi_{\omega}$ = 90$^{\circ}$ and $\phi_{2\omega}$ = 0$^{\circ}$, we observe that the SHG intensity begins to develop below $T_{s2}$, strongly indicating the inversion symmetry being broken in the low-temperature phase.
Indeed, the detailed analysis presented in the following revealed that the observed signal reflects the $\chi^{\rm polar}_{caa}$ component in the $mm2$ point group.

\begin{figure*}[t!]
\centering
\includegraphics[width=12cm]{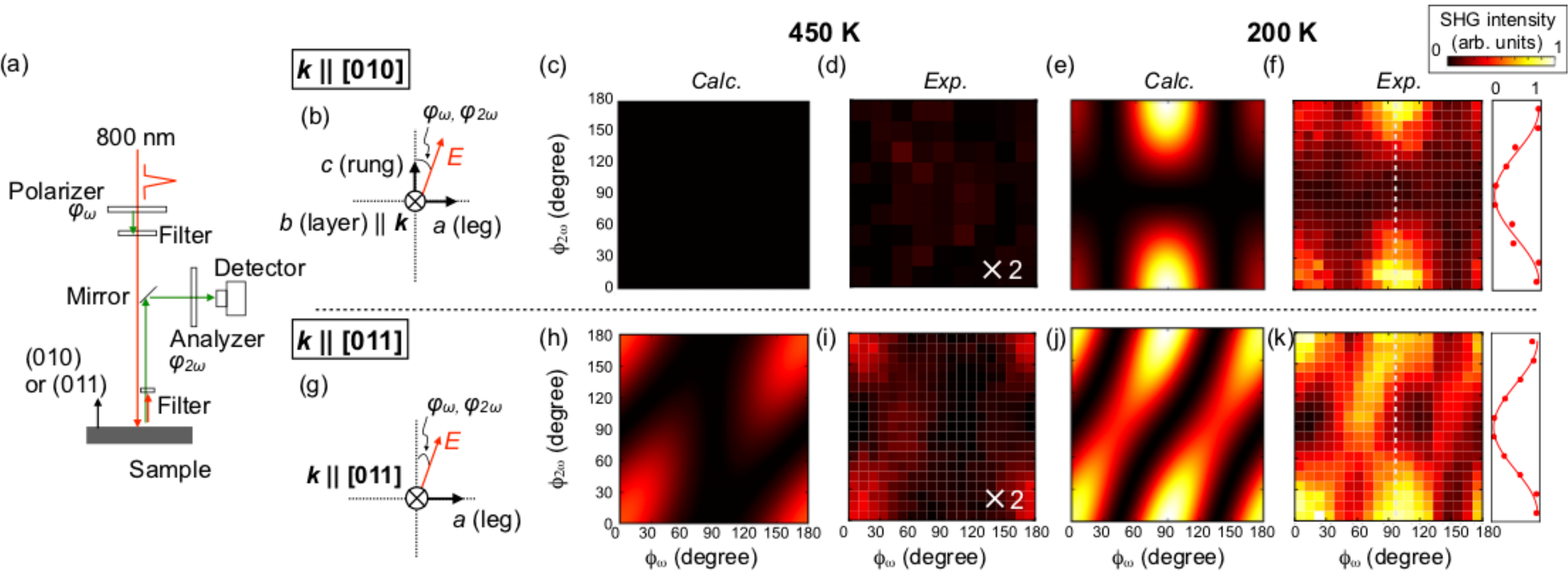}
\caption{Polarization dependence of the SHG intensity for BaFe$_2$Se$_3$. (a) Experimental geometry of SHG measurements. The incident laser beam with the linear polarization is irradiated normal to the (010) or (011) surface, and the reflected SHG is detected by a photomultiplier tube after analyzing the polarization.
(b, g) Relationship between crystal axes and the polarization angle of the incident light $\phi_{\omega}$ and that of SHG $\phi_{2\omega}$. Polarization dependence of (c, e) simulated and (d, f) observed SHG signals at 200 and 450 K for the (010) plane. Polarization dependence of (h, j) simulated and (i, k) observed SHG signals at 200 and 450 K for the (011) plane. In the simulations, the $mmm$ and $mm2$ point groups are respectively assumed for 450 and 200 K data. In the right panel of (f) and (k), $\phi_{2\omega}$ dependences of the SHG intensity collected at $\phi_{\omega}$ = 90$^{\circ}$ are shown.}
\label{Fig3}
\end{figure*}

Here, we discuss the point group below $T_{s2}$ by analyzing the polarization dependence of the SHG signal.
Figures 3(d) and 3(f) show the SHG signal in the $\phi_{\omega}$ - $\phi_{2\omega}$ plane, which are collected in the experimental configuration of Figs. 3(a) and 3(b).
At 450 K, as shown in Fig. 3(d), there is negligibly small SHG signal observed in accordance with the centrosymmetric $Pnma$ symmetry (Fig. 3(c)).
At 200 K, on the other hand, one can see clear polarization dependences: the strong signal is discernible at ($\phi_{\omega}$, $\phi_{2\omega}$) = (0$^{\circ}$, 0$^{\circ}$) and (90$^{\circ}$, 0$^{\circ}$) (Fig. 3(f)).
We here postulate that the point group of the low-temperature phase is $mm2$, which is the subgroup of $mmm$.
The symmetry operation lost in lowering the point group is the inversion operation, so that $mm2$ is a noncentrosymmetric point group.
Then, in $mm2$, $\chi^{\rm polar}_{caa}$, $\chi^{\rm polar}_{ccc}$, and $\chi^{\rm polar}_{aac}$ components can be finite, whereas axial tensors have the same component as in the case of $mmm$ ($\chi^{\rm axial}_{abc}$, $\chi^{\rm axial}_{bca}$, and $\chi^{\rm axial}_{cab}$ can be excited in the present configuration).
Taking care that axial tensors cannot be detected in the current experimental setup, one can simulate the SHG pattern by changing the sign and magnitude of the polar tensor components.
The simulated pattern with $\chi^{\rm polar}_{caa}$ : $\chi^{\rm polar}_{ccc}$ : $\chi^{\rm polar}_{aac}$ = −2.0 : 1.0 : 0 well reproduces the experimentally observed polarization dependence (Fig. 3(e)).

We next performed similar experiments for the (011) surface: the polarization of the initial light ($\phi_{\omega}$) and that of the reflected SHG light ($\phi_{2\omega}$) are defined as shown in Fig. 3(g).
In contrast to the results for the (010) surface, the SHG signals exhibit characteristic polarization dependences even at 450 K (Fig. 3(i)), which is above the second structural transition temperature.
This SHG pattern is well reproduced by putting the axial tensor components in the $mmm$ point group to be $\chi_{bca}^{\rm axial}$ + $\chi_{cab}^{\rm axial}$: $\chi_{abc}^{\rm axial}$  = $-$1.0 : 1.0 (Fig. 3(h)).
We stress here that the axial tensor contributions can be allowed even in centrosymmetric crystals.
When the temperature is decreased down to 200 K, the strong signal appears at ($\phi_{\omega}$, $\phi_{2\omega}$) = (90$^{\circ}$, 0$^{\circ}$) in addition to signals observed at 450 K (Fig. 3(k)).
The observed SHG pattern can be well reproduced by simulations under the assumption of the point group of $mm2$ with the rung direction as the two-fold rotation axis (Fig. 3 (j)); here, we set $\chi^{\rm axial}_{abc}$ : $\chi^{\rm axial}_{bca}$ +$\chi^{\rm axial}_{cab}$ : $\chi^{\rm polar}_{caa}$ : $\chi^{\rm polar}_{ccc}$ : $\chi^{\rm polar}_{cbb}$ + $\chi^{\rm polar}_{bcb}$ = 1 : $-$1 : $-$0.78 : 0.39 : $-$0.67.
We thus conclude that the phase below $T_{s2}$ has the point group of $mm2$ with the polar axis as the rung direction.

Herein, let us identify the crystal structure below $T_{s2}$.
To do this, we first list up possible space groups.
Since any discontinuity or thermal hysteresis is not observed in physical quantities across the transition, the phase transition at $T_{s2}$ is considered to be of the second-order; therefore, we consider the maximal subgroup of $Pnma$.
We then list up three candidate space groups, $Pmc2_1$, $Pmn2_1$, and $Pna2_1$; these respectively lose one of three orthogonal mirror/glide operations associated with the layer, rung, and leg directions.
Among them, $Pmn2_1$ is the most plausible one, because the SHG results indicate that the polar axis is along the rung direction.
We then re-analyze the powder neutron diffraction profiles collected at $T$ = 300 K temperature, yielding better convergence in the $Pmn2_1$ model than in others.
The obtained structural parameters are shown in Table I in the Supplemental Materials~\cite{supple}.
Importantly, the $Pmn2_1$ model is compatible with the block-type magnetic structure below $T_{\rm N}$.
Actually, the powder neutron diffraction profiles below $T_{\rm N}$ can be well fitted by the $Pmn2_1$ structure model and the block-type magnetic structure model with the magnetic wave vector of (1/2, 1/2, 1/2).
Here, the magnetic structure is represented by a single irrep $\Gamma_1$ (Table II of Supplemental Materials~\cite{supple}) in consistent with the second-order nature of the magnetic transition.
The resultant magnetic space group is $Pmn2_11^{\prime}$, so that the time-reversal symmetry is preserved by combining the translation operation even below $T_{\rm N}$ (Fig.1 (d)).
Therefore, the polar-$c$ and axial-$c$ tensors do not become finite upon the magnetic order, justifying the analysis of the SHG results.

The observed polar state in BaFe$_2$Se$_3$ is almost in accordance with the theoretical proposals by Dong, $et$ $al$~\cite{Dong2014}.
According to the theory, the block-type magnetic ordering is stabilized by an electronic origin in a realistic band structures~\cite{Luo2013}.
As a result, there appears uniform displacements of Se atoms perpendicular to the local ladder plane, bearing the macroscopic polarization along the rung direction, which is in consistent with our experimental results.
The microscopic mechanism of this magneto-elastic coupling is argued to be the so-called exchange striction.
The mechanism is known to affect in the up-up-down-down-type antiferromagnetic phase of a conventional multiferroic material $o$-YMnO$_3$ ~\cite{Sergienko2006, Picozzi2007}.
The large lattice distortions induced at around $T_{\rm N}$ shown in ref.~\cite{Nambu2012} strongly support this scenario.

However, there is an apparent incompatibility between the theory and our experimental results.
Whereas the theory predicts emergence of polarization below $T_{\rm N}$, our results indicate that polar lattice distortions emerge below $T_{s2}$, which is much higher than $T_{\rm N}$.
This kind of successive structural and magnetic phase transitions, which are closely related with each other from the symmetry point of view, are reminiscent of the iron-based superconductors including Co-doped BaFe$_2$As$_2$~\cite{Chu2009, Fernandes2014}.
Among them, the high temperature structural transition has been considered to be originating from an orbital order or electronic nematic order, since the rotational symmetry is broken in the electronic system~\cite{Chu2012, Kasahara2012, Fernandes2014, Sun2016}.
It is possible that a similar mechanism is involved in BaFe$_2$Se$_3$, in which prominent quantum fluctuations inherent in the reduced spatial dimensionality results in a large separation of the two transition temperatures.
We also note that the isostructural compound BaFe$_2$S$_3$ exhibits a broad anomaly in the electrical resistivity at around 200 K, which is higher than the antiferromagnetic transition temperature of 120 K.
This could be also due to the possible orbital order~\cite{Yamauchi2015}.
In the iron-based superconductors, not only antiferromagnetic fluctuations but also structural (orbital) fluctuations are expected to have key roles for the formation of superconducting state.
Such possibility shall be also pursued for the superconducting states under pressure in BaFe$_2$S$_3$ and BaFe$_2$Se$_3$, which will open a research field of the superconducting multiferroics.

In conclusion, we observed the optical second harmonic generation signals below the structural transition temperature of 400 K in quasi-one-dimensional ladder material BaFe$_2$Se$_3$.
Combined with neutron diffraction results, we uncovered the detailed crystal structure in the low temperature phase, which has the polar $Pmn2_1$ space group.
In the viewpoint of the symmetry, BaFe$_2$Se$_3$ is therefore multiferroics with both polarity and antiferromagnetic order.
The structural phase transition is driven by the block-type lattice distortions through the magneto striction mechanism.
Considerable change in the electrical resistivity and magnetic susceptibility across the polar-nonpolar phase transition suggests that electronic states are largely modified by this structural instability.

\begin{acknowledgments}
We are grateful to T. Yamauchi, D. Okuyama, F. Du, T. J. Sato, M. Avdeev, and T. Hawai for experimental help and fruitful discussion.
The present work is financially supported by JSPS KAKENHI Nos. 16K17732,
16H01062, 18H04302,
17H05474,
16H04019,
18H01159,
16H04007, 17H05473, 17H06137,
and 17H04844, Research Foundation for Opto-Science and Technology,
Murata Science Foundation,
and Mitsubishi Foundation.
This work was partially performed using facilities of the Institute for Solid State Physics, the University of Tokyo.
\end{acknowledgments}

\bibliography{BaFe2Se3_SHG_paper}

\end{document}


\bibliographystyle{apsrev4-1}
\title{Supplemental materials for \\ "Polar State induced by Block-type Lattice Distortions in BaFe$_2$Se$_3$ with Quasi-One-Dimensional Ladder Structure"}
\author{Takuya Aoyama,
Satoshi Imaizumi,
Takuya Togashi,\\
Yoshifumi Sato,
Kazuki Hashizume,
Yusuke Nambu,
Yasuyuki Hirata,
Masakazu Matsubara, and
Kenya Ohgushi
}
\maketitle

\section{Optical reflectivity spectra}

\begin{figure}[h]
\centering
\includegraphics[width=10cm]{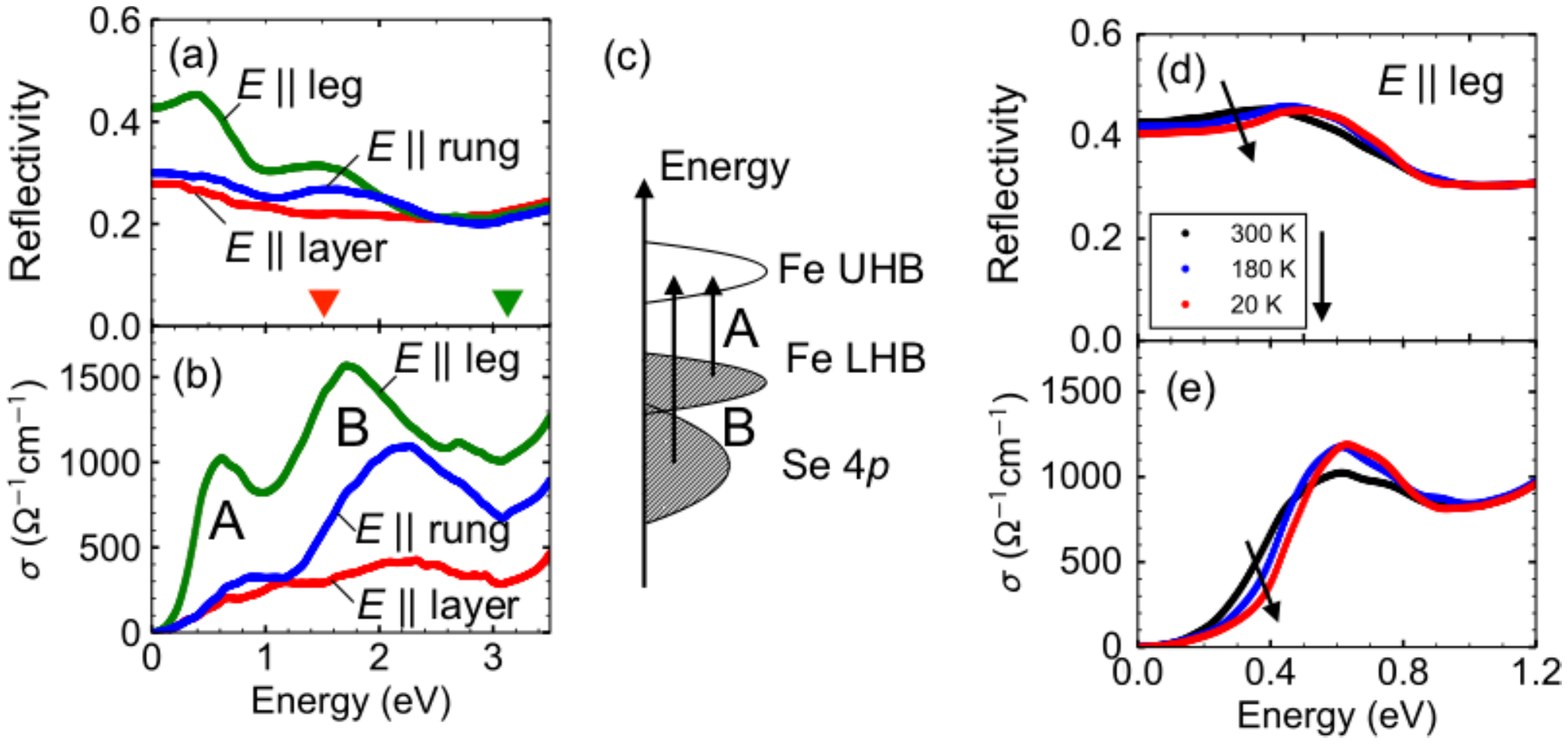}
\caption{Optical reflectivity spectra and optical conductivity ($\sigma$) spectra for BaFe$_2$Se$_3$. (a-b) Polarizaiton dependence of the optical reflectivity and optical conductivity spectra at 300 K. The red and green triangles indicate the wavelength of the incident and emitted lights in second harmonic generation measurements. (c) Schematic band diagrams of BaFe$_2$Se$_3$. LHB and UHB stand for the lower Hubbard and upper Hubbard band, respectively.
(d-e) Temperature dependence of the optical reflectivity and optical conductivity spectra when the light polarization ($E$) is along the leg direction. }
\label{sF1}
\end{figure}

We performed optical reflectivity measurements for the (010) and (001) surfaces of BaFe$_2$Se$_3$, which are polished with Al$_{2}$O$_{3}$ powders.
A Fourier-transform infrared spectrometer and a grating spectrometer were used for measurements in the energy range of 0.1-0.9 and 0.7-4.3 eV, respectively.
The incident light was linearly polarized along three principal axes of the orthorombic crystal structure.
Optical conductivity spectra were obtained from the reflectivity spectra by the Kramers-Kronig transformation.

The optical reflectivity and conductivity spectra taken at 300 K (Figs. S1(a) and S1(b)) show a large polarization dependence.
The optical conductivity spectra ($\sigma$) have large spectral weights, when the light polarization $E$ is  along the leg direction, indicating the quasi-one-dimensional nature of charge dynamics reflecting  the ladder structure.
The key feature is the absence of Drude components in any direction.
This is because the system becomes a Mott insulator due to the prominence of the electron correlation effect in the low-dimensional system.
The optical gap is estimated to be 0.25 eV from the spectra with $E$ along the leg direction, which is consistent with the charge gap deduced from the electrical resistivity measurements.
One can clearly see two optical modes at around 0.6 and 1.7 eV, which are respectively marked as A and B in the figure.
The A mode is considered to be a Mott excitation of Fe-3$d$ orbitals hybridized with Se-4$p$ orbitals.
On the other hand, the B mode is considered to be a charge transfer excitation, which is the transition from the deeper Se-4$p$ orbitals to Fe-3$d$ orbitals.
The schematic picture of transitions is shown in Fig. S1(c).
These optical modes are analogous with that of iron-deficient two-dimensional iron chalcogenide superconductor K$_{2}$Fe$_{4}$Se$_{5}$, which also has block-type magnetism~\cite{Ye_2011, Charnukha2014}.
The SHG experiments described in the main text were performed with the incident energy of 1.55 eV and the emitted energy of 3.10 eV, which are indicated by arrows in Fig. S1. The relevant electronic states in these energy scale are Fe-3$d$ and Se-4$p$ orbitals.

Temperature dependences of reflectivity and optical conductivity spectra with $E$ parallel to the leg direction are shown in Figs. S1(d) and S1(e).
The spectral weights at the low-energy side become smaller with decreasing temperature in accordance with the insulating behavior of the electrical resistivity.

\section{Simulations of second harmonic generation signals}
The polarization dependences of the optical second harmonic generation (SHG) are simulated.
SHG is the second-order nonlinear optical phenomena, in which the incident light with the frequency $\omega$ are converted into the frequency-doubled (2$\omega$) light.
The SHG intensity $I$(2$\omega$) is written by the source term $S$(2$\omega$) as

\begin{eqnarray}
I(2\omega) &\propto& |S(2\omega)|^2 \\
S(2\omega) &=& \mu_0\frac{\partial ^2\vec{P}}{\partial t^2} + \mu_0 (\nabla \times \frac{\partial \vec{M}}{\partial t}) \\
P_i(2\omega) &=& \varepsilon_0\chi^{{\rm polar}}_{ijk} E_j(\omega)E_k(\omega) \\
M_i(2\omega) &=& \varepsilon_0c\chi^{{\rm axial}}_{ijk} E_j(\omega)E_k(\omega)
\end{eqnarray}
The SHG tensors $\chi^{\rm polar}_{ijk}$ and $\chi^{\rm axial}_{ijk}$ depends on the symmetry of the material~\cite{Fiebig05}.
Since the time reversal symmetry is preserved even in the antiferromagnetic state in BaFe$_2$Se$_3$, the SHG tensors should be $i$-tensors.

In $mmm$, which is the point group of the high temperature phase of BaFe$_2$Se$_3$, the axial-$i$ tensor can be finite though polar-$i$ tensor should be zero.
\begin{equation}
  \chi^{\rm axial} = \left[
    \begin{array}{rrrrrr}
               0 & 0 & 0 & \chi^{\rm axial}_{abc} & 0 & 0\\
               0 & 0 & 0 & 0 & \chi^{\rm axial}_{bca} & 0\\
               0 & 0 & 0 & 0 & 0 & \chi^{\rm axial}_{cab}
    \end{array}
  \right]
.
\end{equation}

\begin{equation}
  \chi^{\rm polar} = \left[
    \begin{array}{rrrrrr}
               0 & 0 & 0 & 0 & 0 & 0\\
               0 & 0 & 0 & 0 & 0 & 0\\
               0 & 0 & 0 & 0 & 0 & 0
    \end{array}
  \right]
.
\end{equation}

In $mm2$, which is the most plausible point group of the low temperature phase of BaFe$_2$Se$_3$, several components of the axial-$i$ tensors and polar-$i$ tensors can be finite as follow~\cite{R1963}.
\begin{equation}
  \chi^{{\rm axial}} = \left[
    \begin{array}{rrrrrr}
               0 & 0 & 0 & \chi^{\rm axial}_{abc} & 0 & 0\\
               0 & 0 & 0 & 0 & \chi^{\rm axial}_{bca} & 0\\
               0 & 0 & 0 & 0 & 0 & \chi^{\rm axial}_{cab}
    \end{array}
  \right]
.
\end{equation}

\begin{equation}
  \chi^{{\rm polar}} = \left[
    \begin{array}{rrrrrr}
               0 & 0 & 0 & 0 & \chi^{\rm polar}_{aac} & 0\\
               0 & 0 & 0 & \chi^{\rm polar}_{bbc} & 0 & 0\\
               \chi^{\rm polar}_{caa} & \chi^{\rm polar}_{cbb} & \chi^{\rm polar}_{ccc} & 0 & 0 & 0
    \end{array}
  \right]
.
\label{E_3}
\end{equation}

\begin{figure}[t]
\centering
\includegraphics[width=10cm]{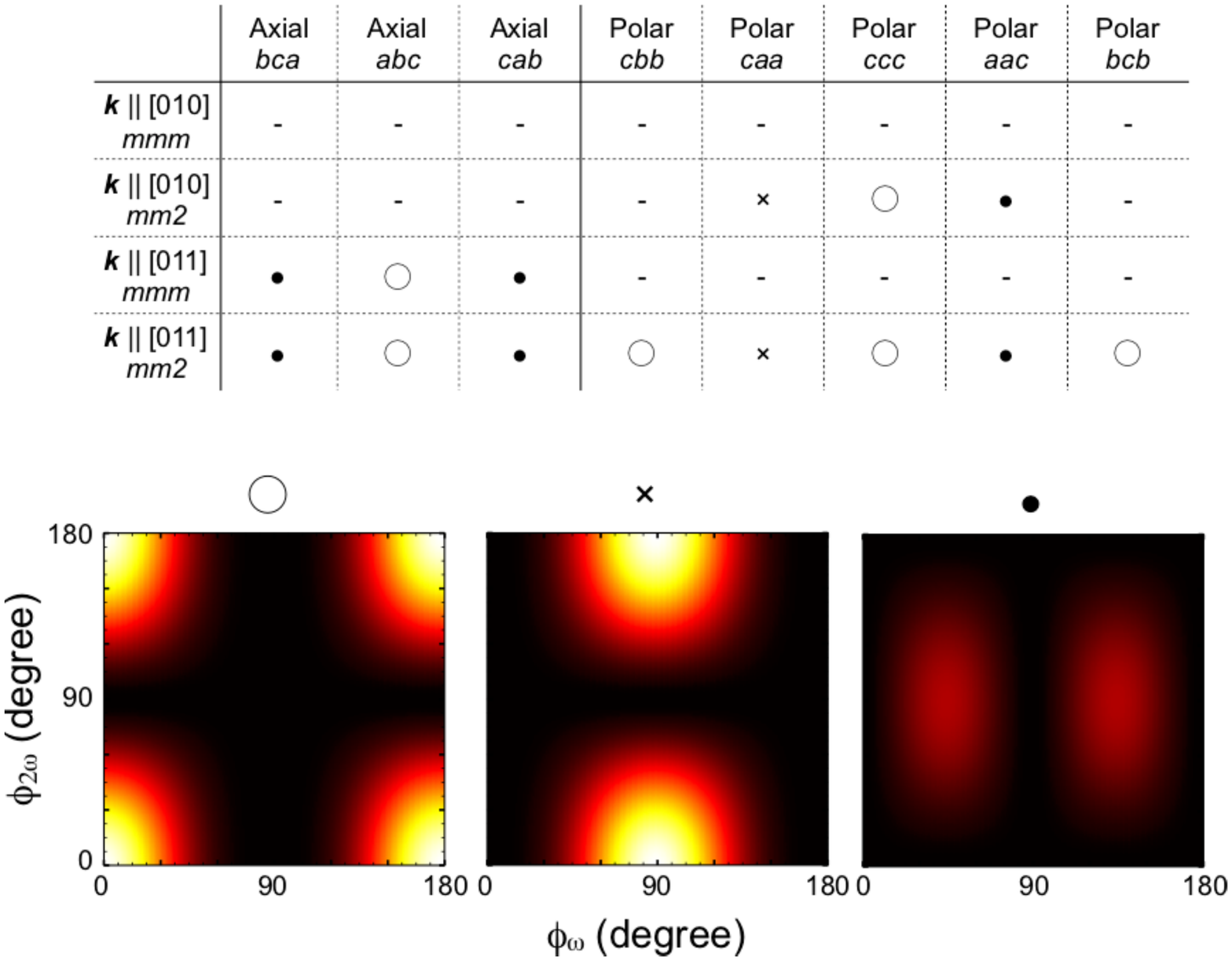}
\caption{Simulated results of the polarization dependences of the optical second harmonic generation for the (010) and (011) surfaces shown in Fig. 3 in the main text. }
\label{sF2}
\end{figure}

Figure S\ref{sF2} shows the calculated results of polarization dependence of the SHG signals with the (010) and (011) configurations shown in Fig. 3 in the main text.
The marks in the table ($\circ$, $\bullet$, $\times$) indicate the corresponding SHG patterns shown in the lower panels.
If several components can be finite, they are expected to interference with each other.

\clearpage

\section{Powder neutron diffraction profiles}
As noted in the main text, the transition across $T_{s2}$ is in the second-order fashion, the space group is therefore assumed to be among the maximal subgroup of $Pnma$.
Three subgroups, $Pmc2_1$, $Pmn2_1$, and $Pna2_1$ out of all subgroups are possible, because they only possess point groups consistent with results inferred from the SHG experiments.
We then tested and evaluated the space group for the data taken at 300 K through the Rietveld refinement on neutron powder diffraction data.
Neutron powder diffraction data were collected on the high-resolution ECHIDNA diffractometer at the Australian Nuclear Science and Technology Organisation (ANSTO)  with $\lambda=2.4395$ {\AA}~\cite{nambu2012}.
The fits yield $R_{wp}=4.40$ for $Pmc2_1$, 4.38 for $Pmn2_1$, and 4.41 for $Pna2_1$.
The difference in the outcomes finds very tiny; however, all the evaluation factors such as $R_p$ and the reduced $\chi^2$ are in the same fashion.
Therefore we concluded the crystal structure below $T_{s2}$ should have the space group $Pmn2_1$ (Table I).

\begin{table}[h!]
\caption{Atomic positions within $Pmn2_1$ of BaFe$_2$Se$_3$ at $T=300$ K determined by Rietveld analysis ($\chi^2=2.45$). Lattice constants are $a=5.43748(4)$ \AA, $b=11.93178(12)$ \AA, and $c=9.16473(9)$ \AA. Isotropic Debye-Waller factor ($B_{\rm iso}$) is employed.}
\label{atomic}
\begin{ruledtabular}
\begin{tabular}{lccccc}
Atom  & Site  & $x$   & $y$   & $z$       & $B_{\rm iso}$ ({\AA}$^2$) \\
\hline
Ba1 & $2a$  & 0 & 0.4397(16)  & 0.7246(21)  & 1.169(150)  \\
Ba2 & $2a$  & 0 & 0.9337(15)  & 0.2582(21)  & 1.169(150)  \\
Fe1 & $4b$  & 0.2393(10)  & 0.7473(7) & 0.8817(30)  & 1.048(38) \\
Fe2 & $4b$  & 0.2405(12)  & 0.7585(6) & 0.5855(30)  & 1.048(38) \\
Se1 & $2a$  & 0 & 0.6034(11)  & 0.9950(27)  & 1.200(41) \\
Se2 & $2a$  & 0 & 0.1159(10)  & 0.9557(25)  & 1.200(41) \\
Se3 & $2a$  & 0 & 0.8780(15)  & 0.7425(31)  & 1.200(41) \\
Se4 & $2a$  & 0 & 0.3779(15)  & 0.2272(28)  & 1.200(41) \\
Se5 & $2a$  & 0 & 0.6566(12)  & 0.4101(26)  & 1.200(41) \\
Se6 & $2a$  & 0 & 0.1445(12)  & 0.5402(25)  & 1.200(41) \\
\end{tabular}
\end{ruledtabular}
\end{table}

Additional magnetic reflections appear in the data below $T_{\rm N}$, and all magnetic-peak positions can be indexed by the magnetic wave vector, $\vec{q}_{\rm m}=(1/2,1/2,1/2)$.
We employed group theoretical analysis to identify the magnetic structure that is allowed by symmetry.
Basis vectors (BVs) of the irreducible representations (irreps) for the wave vector were obtained using the SARA$h$ code~\cite{Wills2000}.
There is only one irrep, and it consists of 12 BVs giving either parallel or antiparallel relationship between uniaxial magnetic moments along one crystallographic axis for two sites out of all the four iron positions per one Wyckoff site (Table II).
To describe magnetic structure, two BVs are required to join to let all the four iron atoms have magnetic moments.
In the case of moments along one crystallographic axis, there are two choices for two combinations, namely four patterns.
Two Wyckoff sites for irons reflect 16 patterns in total.
We assumed that the moment is parallel or antiparallel to the one axis and has the same coefficient of four BVs, and sorted out all 48 patterns to determine the magnetic structure by comparing the $R$-factor.
The best fit with $R_{\rm mag}=5.47$ is the combination of BVs, $\psi_5$ and $\psi_{11}$ for both Fe1 and Fe2, and the second comes $\psi_2$ and $\psi_8$ for $R_{\rm mag}=5.50$.
Both describes magnetic moments perpendicular to the ladder plane.
Note that the cases for moments along the ladder and rung directions poorly describe the data with $R_{\rm mag}>20$.
The optimized refinement of the magnetic structure with $\Phi_{5}$ and $\Phi_{11}$ reproduces the previous result~\cite{nambu2012}, where magnetic moments (2.67(2) $\mu_{\rm B}$/Fe at 5 K) are arranged to form a Fe$_4$ ferromagnetic unit, and it stacks antiferromagnetically along the ladder direction.
The obtained magnetic structure has magnetic point group $mm21^{\prime}$, and belongs to magnetic space group $Pmn2_1 1^{\prime}$ (Fig. S3).

\begin{table}[h!]
\caption{Basis vectors (BVs) of irreducible representations (irreps) for the space group $Pmn2_1$ with the magnetic wave vector $\vec{q}_{\rm m}=(1/2,1/2,1/2)$. Superscripts show the moment direction. Columns for positions represent \#1: $(x,y,z)$, \#2: $(-x+1/2,-y,z+1/2)$, \#3: $(-x,y,z)$, and \#4: $(x+1/2,-y,z+1/2)$ for both Fe1 and Fe2 sites.}
\label{irrep}
\begin{ruledtabular}
\begin{tabular}{cccccccccc}
irrep & BV  & Fe1 (\#1) & Fe1 (\#2) & Fe1 (\#3) & Fe1 (\#4) & Fe2 (\#1) & Fe2 (\#2) & Fe2 (\#3) & Fe2 (\#4) \\
\hline
\multirow{12}{*}{$\Gamma_1$}  & $\psi_1$  & $1^a$ & 0 & $-1^a$  & 0 & $1^a$ & 0 & $-1^a$  & 0 \\
  & $\psi_2$  & $1^b$ & 0 & $1^b$ & 0 & $1^b$ & 0 & $1^b$ & 0 \\
  & $\psi_3$  & $1^c$ & 0 & $1^c$ & 0 & $1^c$ & 0 & $1^c$ & 0 \\
  & $\psi_4$  & 0 & $-1^a$  & 0 & $1^a$ & 0 & $1^a$ & 0 & $-1^a$  \\
  & $\psi_5$  & 0 & $-1^b$  & 0 & $-1^b$  & 0 & $1^b$ & 0 & $1^b$ \\
  & $\psi_6$  & 0 & $1^c$ & 0 & $1^c$ & 0 & $-1^c$  & 0 & $-1^c$  \\
  & $\psi_7$  & 0 & $1^a$ & 0 & $1^a$ & 0 & $-1^a$  & 0 & $-1^a$  \\
  & $\psi_8$  & 0 & $1^b$ & 0 & $-1^b$  & 0 & $-1^b$  & 0 & $1^b$ \\
  & $\psi_9$  & 0 & $-1^c$  & 0 & $1^c$ & 0 & $1^c$ & 0 & $-1^c$  \\
  & $\psi_{10}$ & $1^a$ & 0 & $1^a$ & 0 & $1^a$ & 0 & $1^a$ & 0 \\
  & $\psi_{11}$ & $1^b$ & 0 & $-1^b$  & 0 & $1^b$ & 0 & $-1^b$  & 0 \\
  & $\psi_{12}$ & $1^c$ & 0 & $-1^c$  & 0 & $1^c$ & 0 & $-1^c$  & 0 \\
\end{tabular}
\end{ruledtabular}
\end{table}

\begin{figure}[h!]
\centering
\includegraphics[width=10cm]{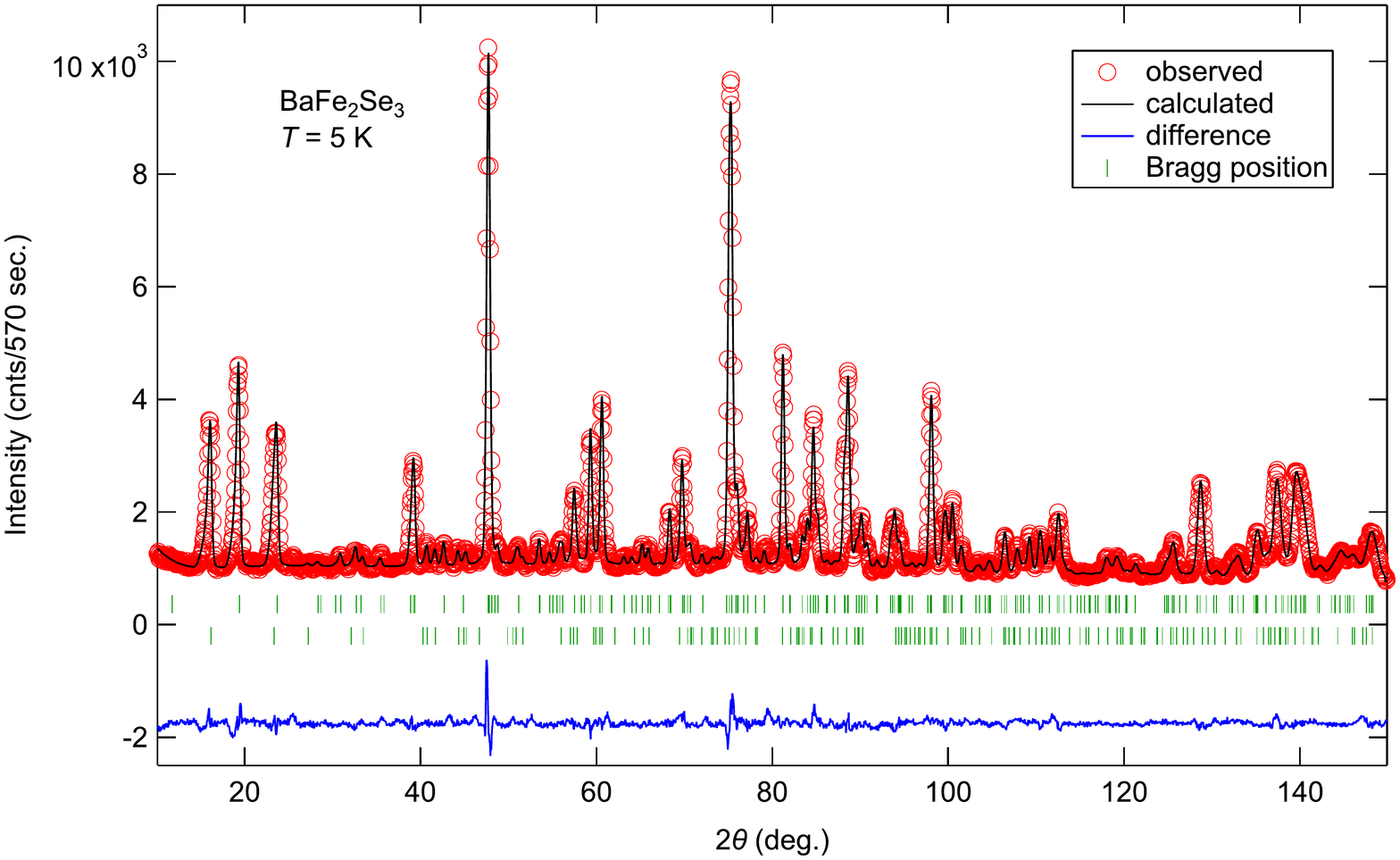}
\caption{High-resolution neutron powder diffraction pattern of BaFe$_2$Se$_3$ taken at 5 K obtained on ECHIDNA with the Rietveld refinement (solid lines). The calculated positions of nuclear and magnetic reflections are indicated (green ticks). The bottom line gives the difference between observed and calculated intensities.}
\label{sF3}
\end{figure}

\clearpage

\bibliography{BaFe2Se3_SHG_paper}